\documentclass[12pt,epsfig]{article}
\usepackage[xdvi]{graphicx}

\newcommand{\beq}{\begin{equation}}
\newcommand{\eeq}{\end{equation}}
\newcommand{\bea}{\begin{eqnarray}}
\newcommand{\eea}{\end{eqnarray}}

\def\e2sig{e^{-2r\sigma}}

\begin{document}
\setlength{\baselineskip}{0.7cm}
\begin{titlepage} 
\begin{flushright}
OCU-PHYS-377
\end{flushright}
\vspace*{10mm}
\begin{center}{\LARGE\bf $H \to Z \gamma$ in Gauge-Higgs Unification}
\end{center}
\vspace*{10mm}
\begin{center}
{\Large Nobuhito Maru}$^{a}$ and {\Large Nobuchika Okada}$^{b}$
\end{center}
\vspace*{0.2cm}
\begin{center}
${}^{a}${\it Department of Mathematics and Physics, Osaka City University, \\
Osaka 558-8585, Japan}
\\[0.2cm]
${}^{b}${\it 
Department of Physics and Astronomy, University of Alabama, \\
Tuscaloosa, Alabama 35487, USA} 
\end{center}
\vspace*{2cm}
\begin{abstract}

In our previous paper~\cite{MO2}, we have investigated effects 
 of a simple gauge-Higgs unification model 
 on the diphoton signal events from the Higgs boson production 
 at the Large Hadron Collider. 
We have found that in this model 
 the effective Higgs-to-diphoton coupling can be 
 enhanced by 1-loop corrections 
 with Kaluza-Klein (KK) modes of bulk fields. 
This result can be an explanation for the observed excess 
 of the signal strength in the Higgs to diphoton decay channel. 
In this paper, we investigate KK-mode effects 
 on another Higgs boson decay mode, $H \to Z \gamma$. 
One naturally expects that the KK modes also 
 contribute to the effective $H-Z-\gamma$ coupling 
 and can cause some deviation for the $H \to Z \gamma$ decay mode 
 from the Standard Model prediction. 
Revealing a correlation between the KK-mode effects 
 on the Higgs-to-diphoton and $H-Z-\gamma$ couplings 
 is an interesting topic in terms of a possibility 
 to discriminate a variety of models for physics 
 beyond the Standard Model. 
We show a very striking result for the gauge-Higgs unification model, 
 namely, the absence of KK-mode contributions to the $H-Z-\gamma$ 
 coupling at the 1-loop level. 
This is a very specific and general prediction 
 of the gauge-Higgs unification scenario.

\end{abstract}
\end{titlepage}

As announced on July 4th 2012, the long-sought Higgs boson 
 was finally discovered by ATLAS \cite{ATLAS} and CMS \cite{CMS} 
 collaborations at the Large Hadron Collider. 
The discovery is based on the Higgs boson search 
 with a variety of Higgs boson decay modes. 
Although the observed data were mostly consistent with 
 the Standard Model (SM) expectations, 
 the diphoton decay mode showed the signal strength 
 considerably larger than the SM prediction. 
Since the effective Higgs-to-diphoton coupling is induced 
 at the quantum level even in the SM, a certain new physics 
 can significantly affect the coupling. 
This fact motivated many recent studies 
 for a possible explanation of the excess 
 in the Higgs to diphoton decay mode by various extensions 
 of the SM with supersymmetry~\cite{AHC-SUSY} or 
 without supersymmetry~\cite{AHC-NonSUSY, MO2}. 
Although the updated CMS analysis~\cite{CMS2} gives 
 a much lower value for the signal strength 
 of the diphoton events than the previous one, 
 the updated ATLAS analysis ~\cite{ATLAS2} is still consistent 
 with their earlier result. 
The excess may persist in future updates.

Gauge-Higgs unification (GHU) \cite{GH} is 
 one of the fascinating scenarios for physics beyond the SM, 
 which can provide us a solution to the gauge 
 hierarchy problem without invoking supersymmetry. 
In this scenario, the SM Higgs doublet is identified 
 with an extra spatial component of a gauge field 
 in higher dimensional gauge theory. 
Nevertheless the scenario is non-renormalizable, 
 the higher dimensional gauge symmetry 
 allows us to predict various finite physical observables 
 such as Higgs potential~\cite{1loopmass, 2loop}, 
 $H\to gg, \gamma\gamma$~\cite{MO, Maru}, 
 the anomalous magnetic moment $g-2$~\cite{g-2}, 
 and the electric dipole moment~\cite{EDM}.

In our previous paper~\cite{MO2}, 
 we have shown that the excess of the Higgs to diphoton decay mode 
 can be explained by 1-loop corrections via Kaluza-Klein (KK) modes 
 in a simple extension of the 5-dimensional minimal GHU model 
 supplemented by color-singlet bulk fermions 
 with a half-periodic boundary condition and appropriately 
 chosen electric charges. 
Through analysis of the renormalization group equation 
 of the Higgs quartic coupling with the so-called 
 gauge-Higgs boundary condition~\cite{GHcondition}, 
 it has been shown that the bulk fermions have also 
 played a crucial role to achieve the observed Higgs boson 
 mass around 125 GeV. 
Since the KK modes have the electroweak charges, 
 we naturally expect that 1-loop diagrams 
 with the KK modes also affect the effective $H-Z-\gamma$ coupling 
 and deviates the Higgs boson partial decay width 
 to $Z \gamma$ from the SM prediction. 
It is an interesting topic to reveal a correlation 
 between the KK-mode effects on the decay modes 
 of $H \to \gamma \gamma$ and $H \to Z\gamma$, 
 because the correlation may show a model-dependent 
 specific property. 
If the excess of the Higgs to diphoton decay mode persists, 
 the correlation between the deviations from the SM predictions 
 for the two decay modes can be a clue to distinguish  
 scenarios beyond the SM, 
 providing a significant improvement of the sensitivity 
 for the Higgs boson signals in the future.

The purpose of this paper is to study the KK-mode contributions 
 to the Higgs to $Z \gamma$ decay in the context of GHU, 
 which has not been addressed in the previous paper~\cite{MO2}. 
Our result is very striking, namely, 
 contributions via the KK modes at the 1-loop level do not exist 
 and hence the effective Higgs to $Z \gamma$ coupling 
 remains unchanged, irrespectively to 
 the effective Higgs to diphoton coupling. 
This result is quite specific for the GHU scenario, 
 which we have never seen in models beyond the SM. 
The reason is the following. 
In the GHU scenario, the electroweak symmetry breaking 
 causes a mass splitting between degenerate KK modes, 
 which one may think analogous to the left-right 
 mixing between sfermions in supersymmetric theories. 
As we will explicitly show below, in the mass eigenstate basis, 
 $Z$ boson always has couplings with two different mass eigenstates 
 corresponding to the splitting mass eigenvalues, 
 while the Higgs boson and photon couple 
 with the same mass eigenstates. 
As a result, there is no 1-loop diagram with the KK modes 
 for the effective Higgs to $Z \gamma$ coupling. 
This specific structure originates from the basic structure 
 of the GHU scenario, where the SM gauge group is embedded 
 in a larger gauge group in extra-dimensions and 
 the SM Higgs doublet is identified with 
 the higher dimensional component of the bulk gauge field.

Now we study the KK-mode contributions to 
 the effective Higgs to $Z \gamma$ coupling. 
In order to show essential points of our discussion, 
 let us consider a toy model of 5-dimensional $SU(3)$ GHU 
 with an orbifold $S^1/Z_2$ compactification for the 5th dimension. 
Although the following discussion is only for this toy model, 
 we expect that our conclusion is applicable to any model 
 of the GHU scenario. 
In the toy model, the $SU(3)$ gauge symmetry is 
 broken to the electroweak gauge group $SU(2) \times U(1)$ 
 by the orbifolding on $S^1/Z_2$ and 
 adopting a non-trivial $Z_2$ parity assignment 
 on 5-dimensional bulk $SU(3)$ gauge field. 
The remaining gauge symmetry $SU(2) \times U(1)$ is 
 supposed to be radiatively broken by the vacuum expectation value 
 of the zero-mode of $A_5$,
 which is identified with the SM Higgs doublet field.

In the toy model, we introduce a bulk fermion 
 of an $SU(3)$ triplet ($\Psi$) and  
 the basic Lagrangian of the model is simply expressed as 
\bea
{\cal L} = -\frac{1}{2} \mbox{Tr}  (F_{MN}F^{MN}) 
+ i\bar{\Psi}D\!\!\!\!/ \Psi, 
\label{lagrangian}
\eea
where the gamma matrix in 5-dimensional theory is defined as 
 $\Gamma^M=(\gamma^\mu, i \gamma^5)$, 
\bea
F_{MN} &=& \partial_M A_N - \partial_N A_M -i g_{5} [A_M, A_N]~(M,N = 0,1,2,3,5), \\
D\!\!\!\!/ &=& \Gamma^M (\partial_M -ig_{5} A_M) \ 
\left( A_{M} = A_{M}^{a} \frac{\lambda^{a}}{2} \ 
\left( \lambda^{a}: \mbox{Gell-Mann matrices} \right) \right),  \\
\Psi &=& (\psi_1, \psi_2, \psi_3)^T.
\eea
Here, the periodic boundary condition along $S^1$ is imposed 
 for all fields\footnote{
 It is possible to impose a half-periodic boundary condition 
 for the bulk fermion. 
 For both cases with the periodic and the half-periodic 
 boundary conditions, the structure of the couplings 
 between the $Z$ boson and KK modes are the same, 
 leading us to the same conclusion. 
Thus, in the following, we explicitly show our results 
 only for the bulk fermion with the periodic boundary condition. 
}.
The non-trivial $Z_2$ parities are assigned 
 for each field using the orbifolding matrix $P={\rm diag}(-,-,+)$, 
\bea 
\label{z2parity} 
A_\mu(-y) = P A_\mu(y) P^{-1}, \quad 
A_5(-y) = -P A_5(y) P^{-1}, \quad 
\Psi(-y) =  P \gamma^5 \Psi(y)
\eea
where $\pm$ means that $Z_2$ parities. 
By these boundary conditions, 
 the $SU(3)$ gauge symmetry is broken to 
 the electroweak gauge group of $SU(2) \times U(1)$.

For the zero-mode of bosonic sector, 
 we obtain exactly what we need for the SM: 
\bea
A^{(0)}_{\mu} &=& \frac{1}{2}
\left(
\begin{array}{ccc}
 W^{3}_{\mu}+ 2 t_W \left(\frac{1}{6} \right) B_{\mu} 
 & \sqrt{2} W^{+}_{\mu} & 0 \\ 
\sqrt{2} W^{-}_{\mu} & 
-W^{3}_{\mu}+ 2 t_W \left(\frac{1}{6} \right) B_{\mu} & 0 \\
 0& 0 & 2 t_W \left(-\frac{1}{3} \right) B_{\mu}   
\end{array}
\right) \nonumber \\
&=& \frac{1}{2}
\left(
\begin{array}{ccc}
\left( c_W - \frac{s_W t_W}{3} \right)Z_\mu +\frac{4}{3} s_W \gamma_\mu 
& \sqrt{2} W^{+}_{\mu}  & 0 \\ 
\sqrt{2} W^{-}_{\mu} & 
-\left( c_W + \frac{s_W t_W}{3} \right)Z_\mu 
-\frac{2}{3} s_W \gamma_\mu & 0 \\ 
0& 0 & 
\frac{s_W t_W}{3} Z_\mu - \frac{2}{3} s_W \gamma_\mu 
\end{array}
\right) \nonumber \\
&=& \frac{1}{2}
\left(
\begin{array}{ccc}
\frac{2}{\sqrt{3}} \gamma_\mu & \sqrt{2} W^{+}_{\mu} & 0 \\
\sqrt{2} W^{-}_{\mu} & -Z_\mu - \frac{1}{\sqrt{3}} \gamma_{\mu} & 0 \\
0& 0 & Z_\mu -\frac{1}{\sqrt{3}} \gamma_\mu 
\end{array}
\right), 
\\ 
A_5^{(0)} &=& \frac{1}{\sqrt{2}}
\left(
\begin{array}{ccc}
0 & 0 & h^{+} \\
0 & 0 & h^{0} \\
h^{-} & h^{0\ast} & 0 
\end{array}
\right), 
\eea 
where $W_\mu^{3}, \ W_{\mu}^{\pm}$ and $B_\mu$ 
 are the $SU(2)$ and $U(1)$ gauge bosons in the SM,  
 $h = (h^{+}, h^{0})^{T}$ is the Higgs doublet field, 
 and $t_W \equiv \tan \theta_W$, 
     $s_W \equiv \sin \theta_W$, 
     $c_W \equiv \cos \theta_W$, 
     with the weak mixing angle $\theta_W$. 
Note that in this toy model 
 the $SU(2) \times U(1)$ gauge groups 
 are unified into a single $SU(3)$ and 
 an unrealistic weak mixing angle, 
 $\sin^{2} \theta_{W} = \frac{3}{4}$, is predicted. 
The zero mode of the bulk fermion is decomposed into 
 an $SU(2)$ doublet left-handed fermion and 
 an $SU(2)$ singlet right-handed fermion. 
Their $U(1)$ charges are the same as those 
 of the quark $SU(2)$ doublet and the $SU(2)$ singlet down-type quark. 
Thus, if the bulk fermion is a QCD color triplet, 
 we may express them as 
$\Psi^{(0)} = 
\left(
u_{L}, d_{L}, d_{R} \\
\right)^T
$. 
However, for our purpose, such an identification of 
 the zero mode fermion with a SM fermion is not important. 
We refer, for example, Ref.~\cite{CCP} 
 for a realistic setup of 5-dimensional GHU scenario.

We are interested in 4-dimensional effective Lagrangian 
 for the KK mode fermions derived from Eq.~(\ref{lagrangian}). 
The bulk field with an even (odd) parity 
 is expanded by the KK mode function $\cos (n y/R)$ ($\sin (n y/R)$) 
 with an integer $n$ and the radius $R$ of the $S^1$. 
Substituting the KK-mode expansion into Eq.~(\ref{lagrangian}) 
 and integrating the KK-mode functions over the fifth dimensional 
 coordinate ($y$), we obtain a 4-dimensional effective Lagrangian. 
The 4-dimensional effective Lagrangian relevant to our discussion 
 are given by 
\bea
{\cal L}_{{\rm fermion}}^{(4D)} 
&\supset&  
\sum_{n=1}^{\infty} \left\{  
i(\bar{\psi}_1^{(n)} ~\bar{\psi}_2^{(n)}  ~\bar{\psi}_3^{(n)}) 
\gamma^\mu \partial_\mu  
\left(
\begin{array}{c}
\psi_1^{(n)} \\
\psi_2^{(n)} \\
\psi_3^{(n)}
\end{array}
\right) \right. \nonumber \\ 
&& \left. 
+\frac{g_4}{2} (\bar{\psi}_1^{(n)}, \bar{\psi}_2^{(n)}, \bar{\psi}_3^{(n)})  
\left(
\begin{array}{ccc}
\frac{2}{\sqrt{3}} \gamma_{\mu} & \sqrt{2} W^{+}_{\mu} & 0 \\
\sqrt{2} W^{-}_{\mu} & -Z_\mu - \frac{1}{\sqrt{3}} \gamma_{\mu}& 0 \\
0& 0 & Z_\mu -\frac{1}{\sqrt{3}} \gamma_\mu
\end{array}
\right) 
\gamma^{\mu} 
\left(
\begin{array}{c}
\psi_1^{(n)} \\
\psi_2^{(n)} \\
\psi_3^{(n)}
\end{array}
\right) \right. \nonumber  \\ 
&& \left. 
- (\bar{\psi}_1^{(n)}, \bar{\psi}_2^{(n)}, \bar{\psi}_3^{(n)})    
\left(
\begin{array}{ccc}
m_{n} & 0 & 0 \\
0 & m_{n} & -m \\
0& -m  & m_{n} 
\end{array}
\right)
\left(
\begin{array}{c} 
\psi_1^{(n)} \\
\psi_2^{(n)} \\
\psi_3^{(n)}
\end{array}
\right) \right\},  
\label{4Deffaction}
\eea
where $\psi_i^{(n)}$ is the $n$-th KK mode Dirac fermion, 
 $m_{n} = \frac{n}{R}$ is the KK mode mass, 
 $g_4 = \frac{g_{5}}{\sqrt{2\pi R}}$ is the 4-dimensional gauge coupling, 
 and $m = \frac{g_4 v}{2} = m_W$ is the mass associated 
 with the electroweak symmetry breaking by 
 the vacuum expectation value of the Higgs doublet ($v$).
In this toy model, $m$ is identical to the $W$-boson mass 
 ($m_W$) because of the unification of the gauge 
 and Yukawa interactions in the GHU scenario. 
In deriving the mass matrix in Eq.~(\ref{4Deffaction}), 
 we have used a chiral rotation 
\beq  
\psi_{1,2,3} \ \to \ e^{-i\frac{\pi}{4}\gamma_{5}} \psi_{1,2,3}  
\eeq
 in order to get rid of $i \gamma_{5}$.

We easily diagonalize the mass matrix for the KK mode fermions 
 by use of the mass eigenstates 
$\tilde{\psi}_{2}^{(n)}, \ \tilde{\psi}_{3}^{(n)}$,   
\bea 
\pmatrix{ 
\psi_{1}^{(n)} \cr 
\tilde{\psi}_{2}^{(n)} \cr 
\tilde{\psi}_{3}^{(n)} \cr 
} 
= U 
\pmatrix{ 
\psi_{1}^{(n)} \cr 
\psi_{2}^{(n)} \cr 
\psi_{3}^{(n)} \cr 
}, \ \ \  
U =\frac{1}{\sqrt{2}}
\left(
\begin{array}{ccc}
\sqrt{2} & 0 & 0 \\
0 & 1 & -1 \\
0 & 1 & 1 
\end{array}
\right), 
\eea
as 
\bea 
U \ 
\left(
\begin{array}{ccc}
m_{n} & 0 & 0 \\
0 & m_{n} & -m  \\
0& -m  & m_{n} 
\end{array}
\right) 
\ U^{\dagger} 
= 
\left(
\begin{array}{ccc}
m_n & 0 & 0 \\
0 & m_n + m & 0 \\
0 & 0 & m_n-m 
\end{array}
\right). 
\eea

In terms of the mass eigenstates for the KK modes, 
 the Lagrangian is described as 
\bea
&&{\cal L}_{{\rm fermion}}^{(4D)}  
\supset \sum_{n=1}^{\infty} 
 \left\{  (\bar{\psi}_1^{(n)}  
  ~\bar{\tilde{\psi}}_2^{(n)}  ~\bar{\tilde{\psi}}_3^{(n)}) 
\right. \nonumber \\ 
&& \left. \times \left(
\begin{array}{ccc}
i \gamma^{\mu} \partial_{\mu} - m_{n} & 0 & 0 \\
0 & i \gamma^{\mu} \partial_{\mu} 
 -\left( m_n + m \right) & 0 \\
0& 0 &i \gamma^{\mu} \partial_{\mu} 
 -\left( m_n - m \right)  
\end{array}
\right)
\left(
\begin{array}{c} 
\psi_1^{(n)} \\
\tilde{\psi}_2^{(n)} \\
\tilde{\psi}_3^{(n)}
\end{array}
\right) \right. \nonumber \\ 
&& \left. 
+\frac{g_4}{2} (\bar{\psi}_1^{(n)}, \bar{\tilde{\psi}}_2^{(n)}, 
\bar{\tilde{\psi}}_3^{(n)})  
\left(
\begin{array}{ccc}
\frac{2 \gamma_{\mu}}{\sqrt{3}} & W^{+}_{\mu} & W^{+}_{\mu} \\
W^{-}_{\mu} & - \frac{1}{\sqrt{3}} \gamma_{\mu} & - Z_\mu  \\
W^{-}_{\mu} & -Z_\mu & - \frac{1}{\sqrt{3}} \gamma_{\mu}
\end{array}
\right) 
\gamma^{\mu} 
\left(
\begin{array}{c}
\psi_1^{(n)} \\
\tilde{\psi}_2^{(n)} \\
\tilde{\psi}_3^{(n)}
\end{array}
\right) \right\}.  
\label{4Deff}
\eea
Now we can see that $Z$ boson ($W$ boson) 
 couples with two different mass eigenstates 
 corresponding to the splitting mass eigenvalues $m_n \pm m$,\footnote{
Because of the coupling of $Z(W)$ boson 
 to fermions with two different mass eigenstates, 
 one might worry about large contributions 
 to electroweak precision measurements, 
 for instance, the KK fermion contributes to the $T$-parameter. 
However, such contributions are suppressed by 
 a factor $(m_W/m_{KK})^2$ and can be safely neglected 
 with a typical KK mode mass scale $m_{KK}$ 
 being of ${\cal O}$(1TeV). 
} 
 while the photon couples to the same mass eigenstates 
 as expected from the electromagnetic $U(1)$ gauge symmetry. 
The coupling between the Higgs boson and the mass eigenstates 
 are found by the replacement $m \to m + \frac{m}{v} H$, 
 where $H$ is the physical Higgs boson. 
Similarly to the photon coupling, 
 the Higgs boson couples to the same mass eigenstates. 
From the structure of the couplings of the KK-mode mass eigenstates 
 with the $Z$ boson, photon and Higgs boson, 
 we conclude that there is no KK fermion contribution 
 to the effective Higgs to $Z \gamma$ coupling 
 at the 1-loop level. 
This result is very typical for the GHU scenario, 
 distinguishable from other models beyond the SM. 
Although calculations are more involved, 
 we can show that the structure of the KK-mode couplings 
 with the $Z$ boson, photon and Higgs boson 
 is the same also for bulk fermions of higher dimensional 
 $SU(3)$ representations.

Our next interest is focused on the contribution  
 by the KK-mode $W$ bosons to the effective Higgs 
 to $Z \gamma$ couplings  
There are two types of interactions of the KK $W$ bosons 
 involving one $Z$ boson. 
One is the 3-point vertex between two KK $W$ bosons and one Z boson, 
 and the other is the 4-point vertex among two KK $W$ bosons, 
 one $Z$ boson and one photon. 
In order to give explicit expressions for the vertices, 
 we first need to find mass eigenstates of the KK $W$ bosons. 
For this purpose, we write a matrix form of the 5-dimensional 
 $SU(3)$ gauge boson as
\bea
A_{\mu} = \frac{1}{2}
\left(
\begin{array}{ccc}
\frac{2}{\sqrt{3}} \gamma_\mu & \sqrt{2} W^{+}_{\mu} & \sqrt{2} A_\mu^{+} \\
\sqrt{2} W^{-}_{\mu} & -Z_\mu - \frac{1}{\sqrt{3}} \gamma_{\mu} & * \\
\sqrt{2}A_\mu^{-} & * & Z_\mu -\frac{1}{\sqrt{3}} \gamma_\mu 
\end{array}
\right),
\eea  
where the $(1,3)$ and $(3,1)$ elements are parity-odd charged 
 gauge boson whose KK modes are mixed with the KK $W$ boson 
 to yield the mass eigenstates. 
In the same way, the $(2,3)$ and $(3,2)$ elements are 
 parity-odd neutral gauge bosons corresponding to 
 the KK modes of $Z$ and photon. 
Since they are irrelevant to our discussion, 
 we have omitted them by the symbol $``*"$ in the matrix.

Using the KK-mode decomposition, 
 we extract the mass terms for the KK $W$ and $A$ gauge bosons 
 from the Lagrangian with the gauge fixing term as 
 (after the electroweak symmetry breaking) 
\bea
&&\int_{-\pi R}^{\pi R} dy 
\left[
-\frac{1}{2} (F_{\mu5}^a)^2 -\frac{1}{2\xi} 
 (\partial^\mu A_\mu^a - \xi \partial^5 A_5^a)^2
\right] \nonumber \\
&\supset& 
-\sum_{n=0}^\infty 
  \left( m_n^2 + m_W^2 \right) W^{\mu+(n)} W_\mu^{-(n)} 
- \sum_{n=1}^\infty 
  \left( m_n^2 + m_W^2 \right) A^{\mu+(n)} A_\mu^{-(n)} \nonumber \\
&& - i 2 m_W 
\sum_{n=1}^\infty  m_n 
\left( W^{\mu+(n)} A_\mu^{-(n)} 
- W^{\mu-(n)} A_\mu^{+(n)} \right). 
\label{mass}
\eea
The mass matrix for the KK modes of 
 $W_\mu^{\pm(n)}$ and $A_\mu^{\pm(n)}$,  
\bea
(W^{\mu-(n)} ~A^{\mu-(n)}) 
\left(
\begin{array}{cc}
m_n^2 + m_W^2 & 2i m_W m_n \\
-2i m_W m_n & m_n^2 + m_W^2 \\
\end{array}
\right)
\left(
\begin{array}{c}
W_\mu^{+(n)} \\
A_\mu^{+(n)} \\
\end{array}
\right),  
\eea
can be easily diagonalized by the following unitary matrix 
\bea
V=
\frac{1}{\sqrt{2}}
\left(
\begin{array}{cc}
1 & i\\
i & 1\\
\end{array}
\right) 
\eea
as
\bea
(P^{\mu-(n)} ~N^{\mu-(n)}) 
\left(
\begin{array}{cc}
(m_n + m_W)^2 & 0 \\
0 & (m_n - m_W)^2 \\
\end{array}
\right)
\left(
\begin{array}{c}
P_\mu^{+(n)} \\
N_\mu^{+(n)} \\
\end{array}
\right). 
\label{massbasis}
\eea
Here, the mass eigenstates are defined as  
\bea
\left(
\begin{array}{c}
P_\mu^{\pm(n)} \\
N_\mu^{\pm(n)} \\
\end{array}
\right)
\equiv
\frac{1}{\sqrt{2}}
\left(
\begin{array}{cc}
1 & i \\
i & 1 \\
\end{array}
\right)
\left(
\begin{array}{c}
W_\mu^{\pm(n)} \\
A_\mu^{\pm(n)} \\
\end{array}
\right)
= \frac{1}{\sqrt{2}}
\left(
\begin{array}{c}
W_\mu^{\pm(n)} + iA_\mu^{\pm(n)} \\
iW_\mu^{\pm(n)} + A_\mu^{\pm(n)} \\
\end{array}
\right). 
\eea
Due to the electroweak symmetry breaking, 
 the KK mode mass eigenvalue $m_n$ splits into $m_n \pm m_W$.

Now we express the 3 point and 4 point vertices 
 in terms of these mass eigenstates.  
For the 3 point vertex, we find 
\bea
&& \int_{-\pi R}^{\pi R} dy {\rm Tr} 
\left[(\partial_\mu A_\nu - \partial_\nu A_\mu)[A_\mu, A_\nu]
\right] \nonumber \\
&\supset& 4 Z^\mu \sum_{n=1}^\infty 
\left[ W_{\mu\nu}^{-(n)}W^{\nu+(n)} - W_{\mu\nu}^{+(n)} W^{\nu-(n)} 
+ A_{\mu\nu}^{+(n)} A^{\nu-(n)} - A_{\mu\nu}^{-(n)} A^{\nu+(n)}
 \right] 
\nonumber \\
&=& 4 i Z^\mu \sum_{n=1}^\infty \left( P_{\mu\nu}^{+(n)} N^{\nu-(n)}
- P_{\mu\nu}^{-(n)} N^{\nu+(n)} + N_{\mu\nu}^{-(n)} P^{\nu+(n)} -
N_{\mu\nu}^{+(n)} P^{\nu-(n)} \right), 
\eea
while for the 4 point vertex, 
\bea
 \int_{-\pi R}^{\pi R} dy {\rm Tr}[A_\mu, A_\nu]^2 &\supset& 
8 \sqrt{3} \sum_{n=1}^\infty
\left[
\gamma^\nu Z^\mu \left( W^{+(n)}_\mu W_\nu^{-(n)} + W^{-(n)}_\mu W_\nu^{+(n)} \right) 
\right. \nonumber \\
&& \left. -4 \gamma^\mu Z_\mu \left( -W^{\nu+(n)} W_\nu^{-(n)} + A^{\nu+(n)} A_\nu^{-(n)} \right) 
\right. \nonumber \\
&& \left. +2 \gamma^\nu Z^\mu \left( A_\mu^{+(n)} A_\nu^{-(n)} + A_\mu^{-(n)} A_\nu^{+(n)} \right)
\right] \nonumber \\
&=& 8 \sqrt{3} i \sum_{n=1}^\infty
\left[
\gamma^\nu Z^\mu 
\left(
- P_\mu^{-(n)} N_\nu^{+(n)} + N_\mu^{-(n)} P_\nu^{+(n)} - P_\mu^{+(n)} N_\nu^{-(n)} \right. \right. \nonumber \\
&& \left. \left. + N_\mu^{+(n)} P_\nu^{-(n)}
\right)
+2 \gamma^\mu Z_\mu (P^{\nu-(n)} N_\nu^{+(n)} - N^{\nu-(n)} P_\nu^{+(n)})
\right].  
\eea
Here, $P_{\mu\nu}^{\pm (n)} \equiv \partial_\mu P_\nu^{\pm(n)} -
 \partial_\nu P_\mu^{\pm(n)}$ and $N_{\mu\nu}^{\pm(n)} \equiv
 \partial_\mu N_\nu^{\pm(n)} - \partial_\nu N_\mu^{\pm(n)}$. 
Therefore, the 3 point and 4 point vertices 
 with one $Z$ boson always involve two different mass eigenstates 
 $P_\mu^{\pm (n)}$ and $N_\mu^{\pm (n)}$ 
 corresponding to the splitting mass eigenvalues, 
 as in the case of the KK fermions.

In the same way, we express the 3 point vertex with one photon 
 and the 4 point vertex with two photons 
 in terms of the mass eigenstates. 
The 3 point vertex is given by 
\bea
&&  \int_{-\pi R}^{\pi R} dy {\rm Tr}
\left[(\partial_\mu A_\nu - \partial_\nu A_\mu)[A_\mu, A_\nu]
\right] \nonumber \\
&\supset& 4 \sqrt{3} \gamma^\mu \sum_{n=1}^\infty \left[ W_{\mu\nu}^{-(n)}W^{\nu+(n)} - W_{\mu\nu}^{+(n)} W^{\nu-(n)} 
+ A_{\mu\nu}^{-(n)} A^{\nu-(n)} - A_{\mu\nu}^{+(n)} A^{\nu+(n)} \right] \nonumber \\
&=& 4 \sqrt{3} \gamma^\mu \sum_{n=1}^\infty \left[ P_{\mu\nu}^{-(n)} P^{\nu+(n)} - P_{\mu\nu}^{+(n)} P^{\nu-(n)} 
+ N_{\mu\nu}^{-(n)} N^{\nu+(n)} - N_{\mu\nu}^{+(n)} N^{\nu-(n)}
\right],  
\eea
while for the 4 point vertex,  
\bea
&& \int_{-\pi R}^{\pi R} dy {\rm Tr}[A_\mu, A_\nu]^2 \supset \nonumber \\ 
&&12 \left[
2\gamma^\mu \gamma_\mu \sum_{n=1}^\infty (W^{\nu+(n)} W_\nu^{-(n)} + A^{\nu+(n)} A_\nu^{-(n)}) \right. \nonumber \\
&& \left. 
+ \gamma^\mu \gamma^\nu \sum_{n=1}^\infty (W_\mu^{+(n)} W_\nu^{-(n)} + W_\mu^{-(n)} W_\nu^{+(n)} + A_\mu^{+(n)} A_\nu^{-(n)} + A_\mu^{-(n)} A_\nu^{+(n)})
\right] \nonumber \\
&=&
12 \left[
2\gamma^\mu \gamma_\mu \sum_{n=1}^\infty (P^{\nu-(n)} P_\nu^{+(n)} + N^{\nu-(n)} N_\nu^{+(n)}) \right. \nonumber \\
&& \left. 
+ \gamma^\mu \gamma^\nu \sum_{n=1}^\infty (P_\mu^{+(n)} P_\nu^{-(n)} + P_\mu^{-(n)} P_\nu^{+(n)} + N_\mu^{+(n)} N_\nu^{-(n)} + N_\mu^{-(n)} N_\nu^{+(n)})
\right].  
\eea
In contrast with the 3 point and 4 point vertices 
 with one $Z$ boson, the resultant vertices 
 involve only the same mass eigenstates.

The 3 point vertex of the mass eigenstates  
 with the Higgs boson is obtained 
 by the replacement $m_W \to m_W + \frac{m_W}{v} H$ 
 in Eq.~(\ref{massbasis}), and hence the structure is 
 the same as the 3 point vertex with one photon. 
Therefore, we arrive at the same conclusion as in the KK fermion case 
 that there is no 1-loop Feynman diagram with the KK gauge bosons 
 for the contribution to the effective Higgs to $Z \gamma$ coupling.

\subsection*{Acknowledgment}
The work of N.M. is supported in part by the Grant-in-Aid 
 for Scientific Research from the Ministry of Education, 
 Science and Culture, Japan No. 24540283.
The work of N.O. is supported in part 
 by the DOE Grant No. DE-FG02-10ER41714.

\
\end{document}